\documentclass[aps,prd,twocolumn,showpacs,nofootinbib]{revtex4}
\usepackage{graphicx} \usepackage{amsmath} \usepackage{amssymb}
\usepackage{amsfonts} \usepackage{bm}

\begin{document}

\newcommand{\be}{\begin{equation}} \newcommand{\ee}{\end{equation}}
\newcommand{\bea}{\begin{eqnarray}}\newcommand{\eea}{\end{eqnarray}}

\title{Boundary contributions to the hypervirial theorem}

\author{J. G. Esteve}  \email{esteve@unizar.es}
\affiliation{Departamento de F\'{\i}sica Te\'orica,
Facultad de Ciencias, Universidad de Zaragoza. Zaragoza, Spain. } 
\affiliation{
Instituto de Biocomputaci\'on y F\'{\i}sica de Sistemas Complejos,
 Universidad de
 Zaragoza. Zaragoza, Spain. }

\author{F. Falceto}  \email{falceto@unizar.es}
\affiliation{Departamento de F\'{\i}sica Te\'orica,
 Facultad de Ciencias, Universidad de Zaragoza. Zaragoza, Spain. } 
\affiliation{
Instituto de Biocomputaci\'on y F\'{\i}sica de Sistemas Complejos,
 Universidad de
 Zaragoza. Zaragoza, Spain. }

\author{Pulak Ranjan Giri} \email{pulak@unizar.es}
\affiliation{Departamento de F\'{\i}sica Te\'orica,
Facultad de Ciencias, Universidad de Zaragoza, Zaragoza, Spain}

\begin{abstract}
It is shown that under certain boundary conditions the 
virial theorem has to be modified. We analyze the origin
of the extra term and compute it in particular examples. 
The Coulomb and harmonic oscillator with point interaction have
been studied in the light of this generalization of the 
virial theorem.
\end{abstract}


\date{\today}

\maketitle

\section{Introduction}

The virial theorem in classical and quantum mechanics has been known 
for a very long time. It appears for the first time in the classical works 
of Clausius in 1870 \cite{Clausius} in the context of statistical mechanics.
See also \cite{Collins} for a historical account.
Already in 1930, Fock \cite{Fock} derived the quantum mechanical version of the
theorem that relates the expectation value of the kinetic energy
in an eigenstate $\psi_n$ of the Hamiltonian to that of the 
Clausius virial function.
More precisely
\begin{eqnarray}\label{virial}
2\langle\psi_n|T\psi_n\rangle=
\langle\psi_n|({\bf x}\cdot\nabla V)\psi_n\rangle.
\end{eqnarray}
The theorem was initially derived for eigenstates
of non relativistic Hamiltonian in flat space
with potentials depending only on the position,
but it can be generalized in many different ways:
for relativistic systems, for a particle moving in an electromagnetic field,  
to many body systems, mixed states  and for compact space
\cite{ini1}-\cite{LiZhangChen} to mention a few. 
Also recently it has been formulated a local version of the virial 
theorem for systems of fermions, see refs. \cite{BrK10} and \cite{BeN10}.

Another kind of generalization is the hypervirial theorem \cite{Hi60} that
works, in principle, for any operator $G$ and can be written
$$\langle\psi_n|[H,G]\psi_n\rangle=0.$$
The virial theorem in dimension $d$ is recovered when 
$$G= -\frac{i}\hbar{\bf x}\cdot{\bf p} - \frac{d}2.$$

In this paper we want to discuss the domain of validity of
these theorems and, for the cases in which they fail, 
we show how to modify them by adding the appropriate
term related to the boundary conditions of the quantum 
system. In this way we extend the applicability of the virial and 
hypervirial theorem to these systems.

In concrete terms, assume that the Hamiltonian is defined over a
domain $D(H)$. $\psi_n$ necessarily belongs to $D(H)$, but it may happen 
that $G \psi_n\not\in D(H)$. In this case the hypervirial theorem
has to be modified by the addition of an extra term $\cal A$:
\begin{eqnarray}\label{virialano}
\langle\psi_n|[H,G]\psi_n\rangle={\cal A},
\end{eqnarray}
with
$${\cal A}\equiv\langle\psi_n| (H_*-E_n)G\psi_n\rangle,$$
where $E_n$ is the energy eigenvalue of $\psi_n$ and
$H_*$ is the closed extension of $H$ to a domain that 
includes $G\psi_n$. Of course, the new term cancels
if $H_*=H$ i.e. when $G\psi_n\in D(H)$, but if that is not the case
it is different from zero in general.
We will see that $\cal A$ can be computed as the 
integration of a pure derivative term and, therefore,
it depends only on the value of the wave function and its derivatives
at the boundary. This fact is reminiscent of the analogous objects 
in quantum field theory like the chiral anomaly \cite{esteve2}. 
We would like to remark that
an additional term for the virial 
(or hypervirial) theorem in classical mechanics may
appear also when the orbits are not bounded in phase space.

This paper is organized in the following fashion: 
in the next section we discuss a simple example to motivate our study
and obtain the generalization of the hypervirial theorem. Then we give two
examples of systems where this generalization is needed, in
section III we discuss the Hydrogen atom problem with point
interaction  in three dimensions while the section IV is devoted to the
isotropic harmonic oscillator. 
Finally we sketch our conclusions in section V. 

\section{Generalized hypervirial theorem}

Before discussing the required generalization
of the hypervirial theorem we would like to
motivate our study by considering a very simple example.

Consider a free particle in one dimension
restricted to move in $[0,\infty)$ and subject to
Robin boundary conditions, i.e. $\psi'(0)+\alpha\psi(0)=0$ with $\alpha>0$.
In this case the free Hamiltonian has a single eigenfunction
$$\psi_0(x)=\sqrt{2\alpha}\, {\rm e}^{-\alpha x},$$
with eigenvalue\footnote{At first sight it may seem strange that $T=p^2/(2m)$
has a negative eigenvalue. Note, however, that $p$ is not symmetric
with Robin boundary conditions and in fact one has
$$\langle\psi |T\psi\rangle=
\frac{\hbar^2}{2m}(\langle\psi' |\psi'\rangle+\psi(0)^*\psi'(0))
=\frac{\hbar^2}{2m}(\langle\psi' |\psi'\rangle-\alpha|\psi(0)|^2),$$
and the second term on the right provides a negative contribution.}
$$E_0=-\frac{\hbar^2\alpha^2}{2m}.$$
If we use (\ref{virial}) to compute the expectation value
of the kinetic energy in this state we obtain that it vanishes,
which is in contradiction with the real result
\begin{equation}\label{expectationT}
\langle\psi_0|T\psi_0\rangle=E_0.
\end{equation}

The reason for this apparent
contradiction is the fact that the domain of the Hamiltonian is not preserved
by the generator of the scale transformation 
\begin{eqnarray}\label{operG}
G= -\frac{i}\hbar  xp - \frac12
\end{eqnarray}
and the virial theorem has to be modified.

To understand the origin of this modification
let us consider in detail
the derivation of the virial theorem. 
We assume a quantum system with infinite dimensional
state space and a self-adjoint
Hamiltonian with domain $D(H)$. If $G\psi_n\in D(H)$ 
we have
\begin{eqnarray}\label{virialder}
\langle\psi_n|[H,G]\psi_n\rangle&=&
\langle\psi_n|HG\psi_n\rangle-
\langle\psi_n|GH\psi_n\rangle =\cr
&=&
\langle H\psi_n|G\psi_n\rangle-
\langle\psi_n|GH\psi_n\rangle = 0
\end{eqnarray}
where in the second line we have used that $H$ is self-adjoint
and that $\psi_n$ is its eigenstate. 

The problems appear if $G\psi_n\not\in D(H)$. In this case the
expressions in the first line of (\ref{virialder})
do not make sense unless we extend $H$ to a larger domain 
that includes $G\psi_n$. In principle it can be done in an arbitrary 
way but in this case there is a well defined prescription
to obtain the required extension. First we restrict $H$ to an 
appropriate dense subspace $S$, typically to functions whose support
does not include the boundary. If we define $\hat H=H|_S$, the
extension $H_*$ is the adjoint of the previous restriction, 
i.e. $H_*=\hat H^+$. We will assume that the restriction is closed
(which could always be achieved by taking the closure of its graph)
and then we know $H_*^+=\hat H$.

With the previous machinery we can rewrite (\ref{virialder})
in a meaningful way
\begin{eqnarray*}
\langle\psi_n|[H_*,G]\psi_n\rangle&=&
\langle\psi_n|H_*G\psi_n\rangle-
\langle\psi_n|GH\psi_n\rangle =\cr
&=&
\langle\psi_n| (H_*-E_n)G\psi_n\rangle\equiv{\cal A},
\end{eqnarray*}
and obtain the generalized virial theorem as announced in the introduction.
Looking at the last line of this expression one could be tempted
to take the adjoint of the operator $H_*$ and make it act in 
the left side of the scalar product. However, this is not possible in
general as the adjoint of $H_*$ is $\hat H$ with a domain smaller than
$H$ that may not contain $\psi_n$. Of course, if it so happens that
$\psi_n\in S$ the additional term cancels.

We would like to show now that under certain assumptions 
(that we actually meet in our examples)
the extra term is a boundary term, i.e. 
it depends only on the value of the functions and their derivatives
at the boundary.

First of all, we rewrite the extra term in a way that is
appropriate for wave functions that are not necessarily
eigenvectors of $H$,
$${\cal A}=\langle\psi_n| H_*G\psi_n\rangle-\langle H \psi_n| G\psi_n\rangle.$$
Then we will show that the value of $\cal A$ does not change if we replace
$\psi_n$ by $\psi_n+\chi$ with $\chi\in S$ and $G\chi\in S$.
The difference reads
\begin{eqnarray}
\Delta{\cal A}&=&
\langle\psi_n| H_*G\chi\rangle-\langle H \psi_n| G\chi\rangle+\cr
&+&
\langle\chi| H_*G\psi_n\rangle-\langle H \chi| G\psi_n\rangle+\cr
&+&
\langle\chi| H_*G\chi\rangle-\langle H \chi| G\chi\rangle.
\end{eqnarray}
Now, given that $G\chi\in D(H)$, we can replace $H_*$ 
by $H$ in the first term of this expression
and, if we take its adjoint, it cancels the second term. For the very 
same reason the last two terms cancel. Also in the third term
we can take the adjoint of $H_*$, because $\chi$ is in the domain of
$H_*^+=\hat H$, and therefore it cancels the fourth term and $\Delta \cal A$ 
vanishes.

Then if, as we supposed before, the elements of $S$ are wave functions 
whose support does not contain the boundary  we deduce that 
the extra term ${\cal A}$ is invariant under deformations of $\psi_n$ that 
do not affect the values of the function and its derivatives at the boundary or,
in other words, the extra term depends only on the latter.

As an illustration to these results and before discussing the
examples of physical interest, we will compute the new term for
the simple system introduced at the beginning of this section.
One immediately gets
\begin{eqnarray}\label{anomaly}
{\cal A}&=&
\frac{\hbar^2}{2m}\int_0^\infty 
\left(G\psi_0(x)\partial_x^2\psi_0^*(x)- 
\psi_0^*(x)\partial_x^2G\psi_0(x)\right) 
{\rm d}x=\cr
&=& 
\frac{\hbar^2}{2m}\int_0^\infty 
\partial_x\big(G\psi_0(x)\partial_x\psi_0^*(x)-\cr 
&&\hskip 1.8cm-\psi_0^*(x)\partial_xG\psi_0(x)\big) 
{\rm d}x
\end{eqnarray}
which is a pure boundary term (for any operator $G$) as stated.

In our particular case it amounts to
\begin{eqnarray}
{\cal A}&=&\frac{\hbar^2}{4m}
\left(3\psi_0(0)\psi_0'(0)^*-\psi_0(0)\psi_0'(0)^*\right)=\cr
&=&-\frac{\hbar^2\alpha}{2m}
\psi_0(0)^2=-\frac{\hbar^2\alpha^2}{m}.
\end{eqnarray}
Now taking into account that
$[G,H]=2T$
and plugging these expressions into the generalized hypervirial 
theorem (\ref{virialano}) we get
$$2\langle\psi_0|T\psi_0\rangle= -\frac{\hbar^2\alpha^2}{m}$$
from which (\ref{expectationT}) follows.

We have seen that if we modify appropriately the hypervirial theorem it 
can be applied to situations in which the action of the operator
$G$ on the energy eigenstate puts it out of the domain of the Hamiltonian.
In the following sections we will apply this idea to the Coulomb
problem and later on to the isotropic harmonic oscillator.

\section{Coulomb potential.}
In this section we shall discuss the virial theorem for the Coulomb 
problem in three dimensions. We shall see that with the chosen boundary 
conditions at the origin the extra term is required to cancel 
the divergence that appears in the application of the virial theorem.

Consider the Hamiltonian
\begin{eqnarray}
H=\frac{{\mathbf p}^2}{2m}- \frac{k}{r}.
\end{eqnarray}
If we separate the angular variables from
the radial one we obtain the Hamiltonian for the latter
\begin{equation}
H_l=\frac{\hbar^2}{2m}\left(
-\frac1r\frac{d^2}{dr^2}r + \frac{l(l+1)}{r^2} - 
\frac{\xi}{r}\right),
\end{equation}
that depends on the angular momentum number $l$.
Here $\xi=2mk/\hbar^2$.

We can simplify the Hamiltonian by
performing a similarity transformation
$\phi=r\psi$. Therefore the new wave
functions $\phi$ are square integrable in $[0,\infty)$
with respect to the Lebesgue measure. 
The new Hamiltonian is then
\begin{equation}
\hat H_l=\frac{\hbar^2}{2m}\left(-\frac{d^2}{dr^2} + \frac{l(l+1)}{r^2} - 
\frac{\xi}{r}\right),
\end{equation}
and it is symmetric with respect to the Lebesgue measure
in $[0,\infty)$ when acting on normalizable wave functions
whose support does not contain the origin.

In order to have a self-adjoint Hamiltonian we have to look for
extensions of $\hat H_l$ with such a property. This issue
has been analyzed in detail in (\cite{alb}). The result
is that if $l\geq 1$ there is a single self-adjoint extension
of $\hat H_l$ whose domain are the wave functions that vanish at $0$,
while $\hat H_0$ has an infinite number of self-adjoint 
extensions characterized by a single parameter $\alpha$.
If we call $H^\alpha$ such a self-adjoint extension, 
its domain is given by
\begin{eqnarray*}
D(H^\alpha)&=&\{\phi\in D(\hat H_0^+)|\, {\rm as}\ r\to 0,\\
&&\hskip 2mm\phi(r)=\phi(0)\left(1-\xi r\ln(|\xi|r)
+\alpha r+o(r)\right)\}.
\end{eqnarray*}
where the little-$o$ notation has been used. 
The case of wave functions that vanish at the origin 
(the standard boundary conditions for the hydrogen atom)
is recovered when  $\alpha\to-\infty$ and $\phi(0)\to0$,
while the product $\phi(0)\alpha$ remains finite.

The point-spectrum of $H^\alpha$ is obtained from the solutions in 
$\lambda$ to the equation
\begin{eqnarray}\label{eqeigen}
F_C(\lambda)\equiv\Psi(1-\lambda)-\ln|\lambda|+\frac1{2\lambda}+2\gamma-1=\frac\alpha\xi,
\end{eqnarray}
such that $\xi/\lambda\geq0$.
Here $\gamma$ is the Euler's constant and
$\Psi$ is the digamma function, i.e. the logarithmic derivative of the 
gamma function. For a solution $\lambda_n$ of the previous equation
the eigenvalue of $H^\alpha$ is 
$$E_n=-\frac{mk^2}{2\hbar^2}\frac1{\lambda_n^2}.$$
In fig. 1 we represent graphically the solutions of 
(\ref{eqeigen}) for $\lambda$, as the intersection
of the branches of the curve with the line
of constant value $\alpha/\xi$. In the negative side
$F(\lambda)$ has an asymptotic value $2\gamma-1\approx0.1544$.

\begin{figure}[h!]
  \centering
    \includegraphics[width=0.45\textwidth]{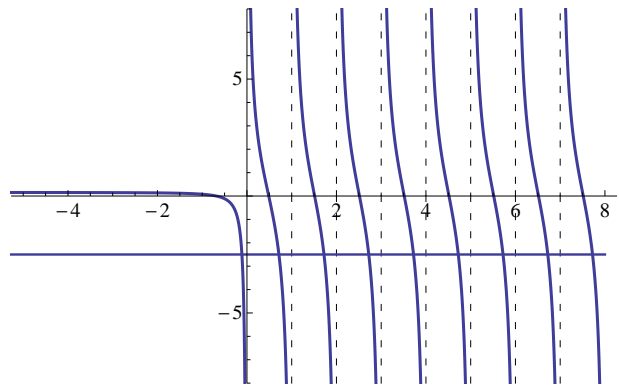}
  \caption{
In the figure we represent $\lambda$
in the horizontal axis and we plot $F_C(\lambda)$.
The point-spectrum of $H^\alpha$ is obtained from the values of
$\lambda$ at the intersection of the graph of $F_C$ with the horizontal 
line at $\alpha/\xi$.
Only those values of $\lambda$ with the same sign that $\xi$ 
are allowed.}
\end{figure}

Note that even in the presence of a repulsive
Coulomb potential, $k < 0$, if we take $\alpha/\xi<2\gamma-1$ 
there is a single bounded state associated to the negative solution
for $\lambda$ of the equation $F_C(\lambda)=\alpha/\xi$. 

The eigenfunction associated to the eigenvalue $E_n$ is given
by the Whittaker function \cite{Abra} conveniently normalized
$$\phi_n(r)=N_n\,W_{\lambda_n,1/2}(\xi r/\lambda_n).$$
With the normalization constant given by
\begin{equation}
N_n^2=\frac{\xi\,|\Gamma(-\lambda_n)|^2}
{2\lambda_n\Psi'(-\lambda_n)+2-\lambda_n^{-1}},
\end{equation}
where the prime denotes derivative.

Now we want to apply the generalized form of the virial theorem
to this system. We take the operator $G=-\frac{i}{2\hbar}  
{\bf x}{\bf p} - \frac34$, then the modified hypervirial theorem reduces to
$$2\langle \psi_n| T\psi_n\rangle+\langle\psi_n | V\psi_n\rangle={\cal A}.$$
Using the energy eigenvalue and integrating out the angular 
coordinates we can write for the radial wave function
\begin{equation}\label{anocoulomb}
{\cal A}-\langle\phi_n|\frac{k}r\phi_n\rangle
=
2E_n
\end{equation}
In this expression we can immediately see the necessity of an 
additional contribution to the virial theorem: the expectation
value of the potential term diverges logarithmically
at the origin (recall that we must use the Lebesgue measure in $[0,\infty)$\ )
while the right hand side is finite.
In fact, we shall see that an analogous divergence appears
in the extra contribution so that both of them cancel to yield
the correct finite result.

We will sketch how to verify it. We first introduce a cut-off $\epsilon$
that removes the origin from the integration region. The extra term only
depends on the boundary conditions of the wave function. 
We can take it from (\ref{anomaly}) and
in its  {\it regularized} version reads
$${\cal A}_\epsilon=-\frac{\hbar^2}{2m}|\phi_n(0)|^2
\left(\xi\ln(|\xi|\epsilon)+\xi-\alpha+\cdots\right),$$
where the dots stand for terms that vanish when $\epsilon\to0$.

We now compute the potential term
$$
\langle\phi_n|\frac{k}r\phi_n\rangle_\epsilon=
N^2k\int_\epsilon^\infty\frac1rW_{\lambda_n,1/2}(\xi r/\lambda_n)^2\ {\rm d} r.
$$
To evaluate the expectation value we use
\begin{eqnarray*}
&&\int_\epsilon^\infty\frac1x W_{\lambda,1/2}(x)^2\ {\rm d} x=\cr
&&=\frac{-1}{\Gamma(-\lambda)^2}
\left(\ln\epsilon-\lambda\Psi'(1-\lambda)
+\Psi(1-\lambda)+2\gamma+\cdots\right),
\end{eqnarray*}
where, as before, the dots represent terms that vanish when 
$\epsilon\to0$. 
{The integral can be obtained using
the identity \cite{PruBryMar} 
\begin{eqnarray}\label{prudnikov}
&&\int \frac1x W_{\mu,\sigma}(x)W_{\rho,\sigma}(x){\rm d}x=\cr
&&\hskip .2cm=
\frac1{\mu-\rho}[W_{\mu,\sigma}(x)W'_{\rho,\sigma}(x)-
W'_{\mu,\sigma}(x)W_{\rho,\sigma}(x)].\end{eqnarray}
} And the series expansion of $W_{\mu,\sigma}$ \cite{Abra}.

Putting everything together and given
$$W_{\lambda,1/2}(0)=1/\Gamma(1-\lambda)$$
and (\ref{eqeigen}) we obtain,
$${\cal A}_\epsilon-\langle\phi_n|\frac{k}r\phi_n\rangle_\epsilon
=
-\frac{mk^2}{\hbar^2}\frac1{\lambda_n^2}+\cdots,
$$
in agreement with (\ref{anocoulomb}).

\section{Harmonic oscillator}

In the previous section we saw that the verification
of the virial theorem were plagued by infinities.
This fact is not inherent to the generalized
virial theorem itself but it is rather related to the
singularity in the Coulomb potential. To illustrate this fact we shall discuss
now a new instance of the generalized virial theorem in three
dimensions that is free of these divergences.

We consider a $3$ dimensional harmonic oscillator given by the 
Hamiltonian
\begin{eqnarray}
H=\frac{{\mathbf p}^2}{2m}+ \frac{1}{2}m\omega^2r^2\,,
\end{eqnarray}
As before, we factorize the angular part and 
study the radial Hamiltonian for the radial part
$\psi(r)$ of the wave function of angular momentum number $l$:
$$H_l=\frac{\hbar^2}{2m}\left(
-\frac1r\frac{d^2}{dr^2}r + \frac{l(l+1)}{r^2}\right) 
+ \frac12 m\omega^2r^2.
$$
Or acting on $\phi(r)=r\psi(r)$
$$
\hat H_l=\frac{\hbar^2}{2m}\left(-\frac{d^2}{dr^2} + \frac{l(l+1)}{r^2}
\right)
+ \frac12 m\omega^2r^2,
$$
that is a symmetric operator with respect to the Lebesgue measure 
in $[0,\infty)$ when acting on functions whose support does not 
include the point 0.

In order to obtain a self-adjoint operator we have to extend
$\hat H_l$ to a larger domain. As it was the case in the previous
section if $l\geq 1$ the only self-adjoint extension is to functions
that vanish at the origin. In this case the extra term cancels and
the standard virial theorem holds. However, if $l=0$, we have a
whole family of self-adjoint extensions parametrized by $\beta$.
If we call $H^\beta$ such an extension its domain corresponds 
to Robin boundary conditions, i.e.
\begin{eqnarray}\label{domain2}
D(H^\beta)&=&\{\phi\in D(\hat H_0^+)| \phi'(0)+2\beta\phi(0)=0\}.
\end{eqnarray}

The normalized eigenfunctions of $H^\beta$
can be written in terms of the Whittaker function as
\begin{eqnarray}
\phi_E(r)= \frac{N}{\sqrt{\sigma r}}
W_{\kappa,1/4}(\sigma^2r^2)
\,,\label{eigen3}
\end{eqnarray}
where $\sigma= (m \omega/\hbar)^{1/2}$, $\kappa=\frac{E}{2\hbar\omega}$ and the
normalization constant $N$ is obtained from the condition
$$\frac{N^2}{2\sigma}\int_0^\infty\frac1xW_{\kappa,1/4}(x)^2{\rm d}x=1.$$
The integral can be performed using (\ref{prudnikov})
and finally we get
$$N^2=
\frac{2\sigma}\pi
\frac{\Gamma(3/4-\kappa)\Gamma(1/4-\kappa)}{\Psi(3/4-\kappa)-\Psi(1/4-\kappa)}.$$

Only for specific values of $\kappa$ the eigenstates  (\ref{eigen3})  belong
to  domain (\ref{domain2}), demanding $\phi_E \in D(H^\beta)$ we 
obtain the equation for $\kappa$
\begin{eqnarray}\label{eigenharm}
F_H(\kappa)\equiv\frac{\Gamma(3/4- \kappa)}{\Gamma(1/4- \kappa)}= \frac\beta\sigma
\,.
\end{eqnarray}
The solutions are represented in fig. 2.

\begin{figure}[h!]
  \centering
    \includegraphics[width=0.45\textwidth]{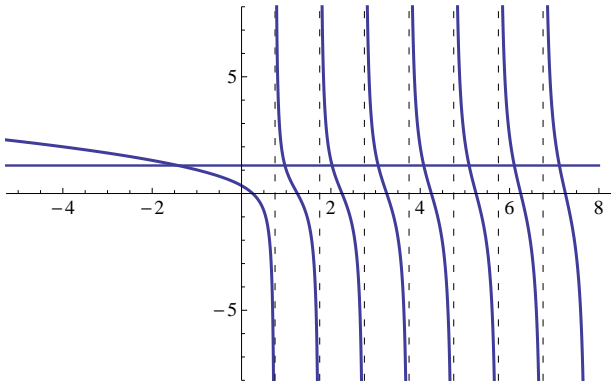}
  \caption{
In the figure we represent $F_H(\kappa)$
with $\kappa$ in the horizontal axis. 
The point spectrum of $H^\beta$ are obtained from the values of
$\kappa$ at the intersection of the graph of $F_H$ with the horizontal 
line at $\beta/\sigma$.
The possible values of $\kappa$ are given as the 
intersection of the curve with the horizontal line at $\beta/\sigma$.}
\end{figure}

In general the eigenvalues $E_n=2\kappa_n \hbar \omega$ must be computed 
numerically. One particular case in which we have analytic solutions 
is $\beta=0$, that correspond to Neumann
boundary conditions. In this case we get 
$E_n= n\hbar \omega+1/2$.
We also have exact results for $\beta=-\infty$, Dirichlet boundary conditions
and $E_n= n\hbar \omega+3/2$.
The latter corresponds to 
the harmonic oscillator without point interaction.
Note also that for $\beta/\sigma>\Gamma(3/4)/\Gamma(1/4)\approx0.337989...$, 
there is an eigenstate with negative energy. Of course, as discussed in the 
footnote of section II, there is not a contradiction between the negative 
eigenvalue and the fact that $H^\beta$ is the sum of squares of operators.
The point is that with Robin boundary conditions some of those operators
are not self-adjoint and there is not any reason to argue that $H^\beta$ is 
semi-positive definite. The same applies to the Coulomb potential with 
repulsive interaction that we discussed in the previous section. 

The generalized virial theorem, in terms of the radial wave function 
$\phi_n$,
reads
$$\langle\phi_n|T\phi_n\rangle -\langle\phi_n|V\phi_n\rangle =\cal A$$
were we have taken $G$ as before.
Using the energy eigenvalue we can rewrite the identity above 
in the following way
\begin{equation}\label{anovirharmonic}
2\langle\phi_n|V\phi_n\rangle 
+ {\cal A} =E_n,
\end{equation}
and this is the expression that we will check.

We first compute the additional term
\begin{eqnarray}\label{anoharmonic}
{\cal A}&=&\frac{\hbar^2}{4m}
\phi_n(0)\phi_n'(0)^*
=-\frac{\hbar^2\beta}{2 m}
|\phi_n(0)|^2=\cr
&=&-N^2\frac{\pi\sigma\hbar^2}{2
m}
\frac1{\Gamma(3/4-\kappa_n)\Gamma(1/4-\kappa_n)}.
\end{eqnarray}
In the last equality condition (\ref{eigenharm}) was used.

The expectation value of the potential can be written
\begin{eqnarray}
\langle\phi_n|V\phi_n\rangle =
N^2\frac{\sigma\hbar^2}{4m}\int_0^\infty W_{\kappa_n,1/4}(z)^2{\rm d} z,
\end{eqnarray}
where the change of variable $z=\sigma^2 x^2$ was performed.

Now we can compute the previous integral with the help of the identity
\begin{eqnarray}
zW_{\kappa,1/4}(z)&=&
W_{\kappa+1,1/4}(z)+
2\kappa W_{\kappa,1/4}(z)+\cr
&&+(3/4-\kappa)(1/4-\kappa)W_{\kappa-1,1/4}(z),
\end{eqnarray}
that we apply to one of the factors of the square inside 
the integral.
We perform the resulting integrals using (\ref{prudnikov})
and finally we obtain
\begin{eqnarray}
\langle\phi_n|V\phi_n\rangle &=&
\frac{\sigma^2\hbar^2\kappa_n}{m}
+\cr
&&+N^2\frac{\pi\sigma\hbar^2}{4m}
\frac1{\Gamma(3/4-\kappa_n)\Gamma(1/4-\kappa_n)}.
\end{eqnarray}
If we plug this expression and (\ref{anoharmonic}) into 
(\ref{anovirharmonic}) we can see that the identity holds. 

\section{Conclusion}
We proved that, when the dilations symmetry 
does not keep invariant the domain of 
definition of the Hamiltonian, the exact virial theorem has an extra term 
that accounts for this fact. 
We have shown that this term is the integral of a pure derivative 
and, therefore, its value depends only of the behaviour of the wave function
at the boundary.
We have verified this generalization of the 
virial theorem in the cases of Coulomb and harmonic oscillator with 
point interaction in three dimensions.

As an extension of our work, it could be interesting to study
if a similar phenomenon occurs for the local  virial theorem 
for fermions with harmonic potential \cite{BrK10}\cite{BeN10},
and the boundary conditions used in this paper.

\section{Acknowledgment}
This work is supported by CICYT (grant FPA2009-09638) and DGIID-DGA (grant 2010-E24/2).

\end{document}